\documentclass[a4paper,twocolumn,11pt,accepted=2025-07-03]{quantumarticle}
\pdfoutput=1
\usepackage[utf8]{inputenc}
\usepackage[english]{babel}
\usepackage[T1]{fontenc}
\usepackage{amssymb,amsthm,mathtools,dsfont}
\usepackage{physics}
\usepackage[colorlinks,citecolor=myBlue,linkcolor=magenta,urlcolor=myBlue]{hyperref}  
\usepackage[capitalise]{cleveref}
\usepackage{ulem}

\usepackage{tikz}
\usepackage{lipsum}
\usepackage{orcidlink}

\crefname{appendix}{App.}{Apps.}
\crefname{section}{Sect.}{Sects.}

\DeclareMathOperator{\1}{\mathds{1}}

\DeclareMathOperator{\Nd}{\mathds{N}}
\DeclareMathOperator{\e}{e}
\DeclareMathOperator{\AAA}{\mathcal{A}}

\DeclareMathOperator{\phat}{\mathit{\hat{p}}}
\DeclareMathOperator{\qhat}{\mathit{\hat{q}}}
\definecolor{myBlue}{RGB}{3,3,133} 
\definecolor{ngreen}{rgb}{0.3,0.7,0.3}
\definecolor{nblue}{rgb}{0.3,0.3,1.0}
\definecolor{ngreen}{rgb}{0.2,0.7,0.2}
\definecolor{nred}{rgb}{0.9,0.1,0}
\definecolor{nblack}{rgb}{0,0,0}
\definecolor{qpurple}{HTML}{53257f}

\begin{document}

\title{Graphical Framework for Non-Gaussian Quantum States}

\author{\orcidlink{0000-0003-2753-6027} Lina Vandr\'e}
\address{State Key Laboratory for Mesoscopic Physics, School of Physics, Frontiers Science Center for Nano-optoelectronics, Peking University, Beijing 100871, China}
\address{Naturwissenschaftlich-Technische Fakult\"{a}t, Universit\"{a}t Siegen, Walter-Flex-Stra{\ss}e 3, 57068 Siegen, Germany}
\address{Technische Universität Wien, Atominstitut, Vienna Center for Quantum Science and Technology, Stadionallee 2, 1020 Vienna, Austria}
\thanks{Equal contributions.}

\author{\orcidlink{0000-0002-9303-604X} Boxuan Jing}
\address{State Key Laboratory for Mesoscopic Physics, School of Physics, Frontiers Science Center for Nano-optoelectronics, Peking University, Beijing 100871, China}
\thanks{Equal contributions.}

\author{\orcidlink{0000-0002-8584-7985} Yu~Xiang}
\email{xiangy.phy@pku.edu.cn}
\address{State Key Laboratory for Mesoscopic Physics, School of Physics, Frontiers Science Center for Nano-optoelectronics, Peking University, Beijing 100871, China}
\affiliation{Ministry of Education Key Laboratory for Nonequilibrium Synthesis and Modulation of Condensed Matter, Shaanxi Province Key Laboratory of Quantum Information and Quantum Optoelectronic Devices, School of Physics, Xi'an Jiaotong University, Xi'an 710049, China}

\author{\orcidlink{0000-0002-6033-0867} Otfried G\"uhne}
\address{Naturwissenschaftlich-Technische Fakult\"{a}t, Universit\"{a}t Siegen, Walter-Flex-Stra{\ss}e 3, 57068 Siegen, Germany}

\author{\orcidlink{0000-0002-2408-4320} Qiongyi~He}
\email{qiongyihe@pku.edu.cn}
\address{State Key Laboratory for Mesoscopic Physics, School of Physics, Frontiers Science Center for Nano-optoelectronics, Peking University, Beijing 100871, China}
\address{Collaborative Innovation Center of Extreme Optics, Shanxi University, Taiyuan, Shanxi 030006, China}
\address{Hefei National Laboratory, Hefei 230088, China}

\begin{abstract}
We provide a graphical method to describe and analyze non-Gaussian quantum states using a hypergraph framework. These states are pivotal resources for quantum computing, communication, and metrology, but their characterization is hindered by their complex high-order correlations. The  framework encapsulates  transformation rules for a series of typical Gaussian unitary operation and local quadrature measurement, offering a visually intuitive tool for manipulating such states through experimentally feasible pathways. Notably, we develop methods for the generation of complex hypergraph states with more or higher-order hyperedges from simple structures through Gaussian operations only, facilitated by our graphical rules. We present illustrative examples on the preparation of non-Gaussian states rooted in these graph-based formalisms, revealing their potential to advance continuous-variable general quantum computing capabilities.
\end{abstract}
\maketitle

\section{Introduction} \label{sec:intro}

\begin{figure}
    \centering
\includegraphics[width=0.4\textwidth]{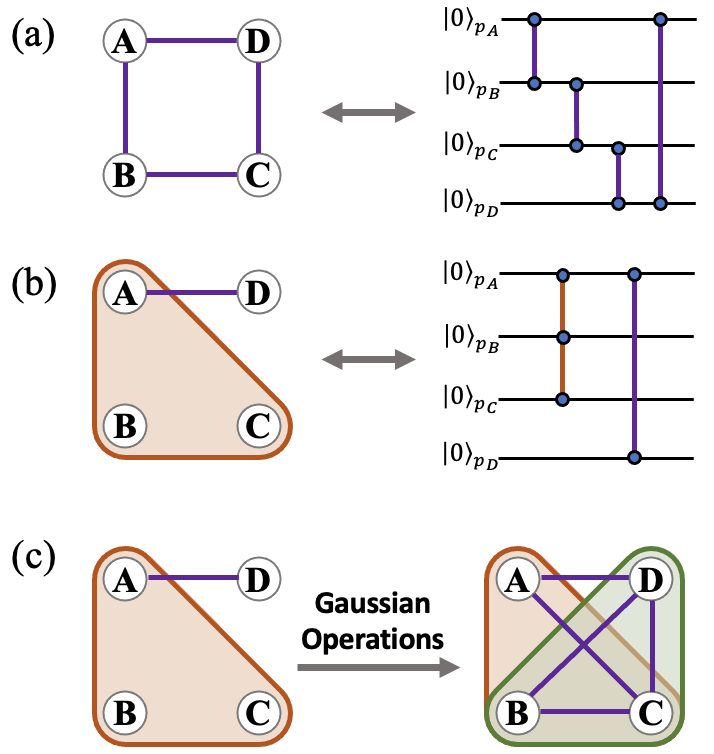}
    \caption{Examples of graph  and hypergraph states, as well as the graphical framework for hypergraph states as discussed in this paper. 
    (a) Topology of a four-mode square graph state and the circuit to generate it. 
    (b) A four-mode hypergraph state. Hypergraph states generalize graph states by allowing hyperedges (colored closed shapes), which represent interactions between multiple modes.
    (c) Based on the graphical rules proposed in this paper, a complex hypergraph state with more or higher-order hyperedges can be generated from simple structures solely through Gaussian operations. 
    }
    \label{fig:simple_hgraph}
\end{figure}

In the quest to demonstrate a genuine quantum advantage in the continuous 
variable (CV) domain, substantial progress on the controllable generation 
of multimode non-Gaussian states has recently been made by photon 
subtraction 
\cite{ra2020non,stiesdal2021controlled,PhysRevX.7.031012}
 and nonlinear unitary operations such as spontaneous parametric down-conversion (SPDC) \cite{Chang20spntaneousdownconv}. 
These strategies usually introduce higher-order statistical moments of the quadrature field operators, go beyond Gaussian states and therefore have irreplaceable advantages as quantum resource \cite{Zhuang18resource,PhysRevX.8.041038} 
in various tasks, including entanglement
distillation \cite{Eisert02distillation,Fiurasek02distillation}, quantum metrology \cite{colombo2022time}, and CV quantum computation \cite{Niset09errorcor,Lloyd99quantumcomp,Menicucci06clusterquantumcomp,PhysRevLett.130.090602}.
Compared to the remarkable progress in experimental realizations, how to understand and more definitely classify the multimode non-Gaussian states with high-order interplay among quadratures, still faces a formidable challenge. 
To address it in a more intuitive and systematic manner, graphical frameworks \cite{graphtheopy1,graphtheopy2,arxiv2012,Duncan_2020_ZX,perezgarcia2007matrixproductstaterepresentations,Bridgeman_2017_TensorNetworks,lockhart2016combinatorialentanglement, PentBI2013, CSW2014, MultGraphAppr2014, Vandre2022, feynman2018space,hasselmann1966feynman, veltman1994feynman} 
as means of describing quantum states emerge as a highly promising solutions.

Graph states \cite{Hein_2004,thomas2022efficient,hein2006entanglement} and hypergraph states \cite{Kruszynska_2009, Qu_2013, Rossi_2013, guhne2014entanglement, jwwnc} were first defined in discrete variable (DV) 
systems to form multiparticle entangled states corresponding to mathematical graphs. They have applications in various contexts, ranging from quantum error correction \cite{Shor_1995, Wagner_2018} to measurement-based quantum computation \cite{Raussendorf_2001,Gachechiladze_2019}. This kind of definition in terms of graphical formalism then got generalised to CV system \cite{Zhang08graphical,zhang06clusterstates}. Examples are shown in Figs.\ \ref{fig:simple_hgraph}(a) and (b).
It has been proven that any Gaussian state can be described by a CV graph state \cite{Menicucci11graphrules}. The first and second-order statistical moments of the quadratures,
which completely characterise a Gaussian state,
can be derived from the edges in the corresponding graph. Naturally, CV hypergraph states have richer graphical structures \cite{Moore19CVhyper,Takeuchi_2019_CVhypergraph},
since they include multimode interactions. 
Such inherent multimode nonlinearity can represent the high-order correlation in non-Gaussian states. 
Ref.~\cite{Moore19CVhyper} showed an evidence that a 3-order hypergraph state restricted to a Gaussian measurement strategy can be a candidate for universal quantum computing. Nevertheless, the ramification of arbitrary Gaussian operations on non-Gaussian hypergraph states remains an underexplored frontier, given their potential to yield higher-order hypergraph states that distribute non-Gaussian resources across an expanded set of modes.

Here, we provide graphical rules which serve  as a comprehensive framework for manipulating non-Gaussian states in CV systems. This pictorial representation encapsulates hypergraph transformation rules encompassing typical Gaussian unitary operations, together with local quadrature measurements, offering an intuitive visual modality for implementing experimentally feasible operations on CV non-Gaussian states.
Crucially, as shown in \cref{fig:simple_hgraph}(c), we demonstrate the capability to generate complex hypergraph states with more or higher-order hyperedges from simple structures solely through Gaussian operations, underscoring the power and versatility of the graphical rules provided here. 
Furthermore, we give three examples of non-Gaussian state preparation schemes rooted in our graph-based formalism. 
These instances illustrate the practical potential of the graphical framework, positioning it as a valuable tool for advancing the frontiers of quantum information processing.

\section{CV Hypergraph States} 
A weighted hypergraph is a tuple $H = (V,E,T)$ with a set $V=\{v_i\}_{i=1}^{n}$ of $n$ vertices, 
a set $E=\{e_j\}_{j=1}^{|E|}$ of ${|E|}$ hyperedges, and a set $T=\{t_{e_j}\}_{j=1}^{|E|}$ 
of real valued weights associated to hyperedges, where $|E|$ denotes the number of elements 
of $E$. 
A CV hypergraph state $\ket{H}$ is defined  as an $n$-mode quantum state where the vertices represent the modes of the state and the weighted hyperedges represent generalized controlled-$Z$ gates performed on the modes \cite{Takeuchi_2019_CVhypergraph,Moore19CVhyper}. More formally, we have 
\begin{align}
    \ket{H} = \prod_{e_j \in E} C_{e_j}(t_{e_j}) \ket{0}_p^{\otimes 
n},
\end{align}
where $\ket{0}_p$ is the zero-momentum eigenstate in the limit of infinite squeezing with normalization $\bra{p'}\ket{p}=\delta_{p,p'}$ 
and $C_{e_j}(t_{e_j})= \e^{i t_{e_j} \qhat_{v_1} \qhat_{v_2} \dots \qhat_{v_m}}$ is a 
generalized controlled-$Z$ gate acting on $m$ modes connected by the $j-$th hyperedge 
$e_j = \lbrace v_1, v_2, \dots v_m \rbrace$. 
If a hyperedge only contains two vertices, $C_e$ is the standard controlled-$Z$ gate $C_e(t) = \e^{i t \qhat_{v_1} \qhat_{v_2}}$ and if it contains three vertices, it is the Toffoli gate $C_e(t) = \e^{i t \qhat_{v_1} \qhat_{v_2} \qhat_{v_3}}$ 
\cite{PhysRevResearch.6.023332}. We use the shorthand notation $q_{e_j} \coloneqq \qhat_{v_1} \qhat_{v_2} \dots \qhat_{v_m}$
and define the quadrature operators in units such that $[\qhat, \phat] = i$ holds.
 A hypergraph state of $k$-th order is classified by the maximum number of $k$ modes connected with one 
hyperedge.

\begin{figure}
    \centering
\includegraphics[width=0.47\textwidth]{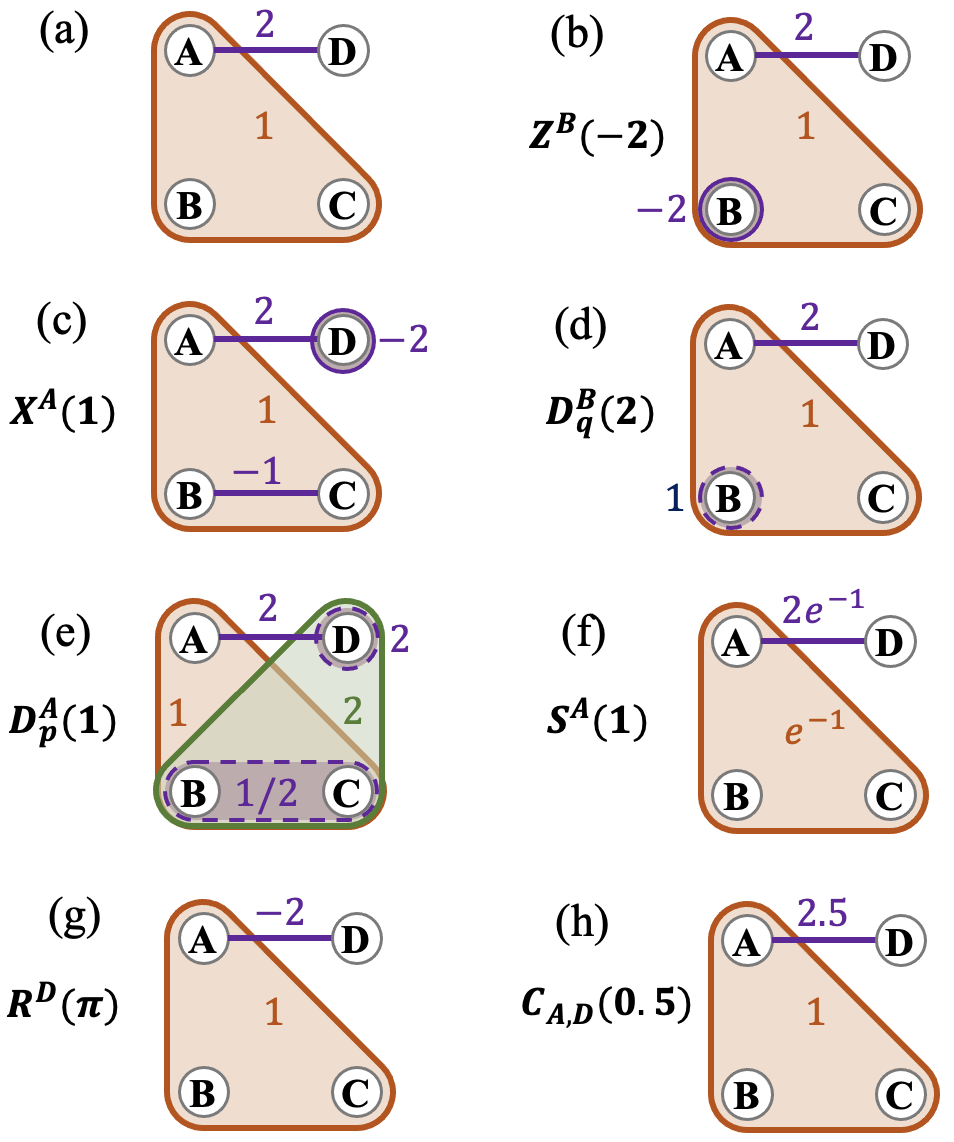}
    \caption{
Examples of a weighted hypergraph state and different Gaussian operations applied on it. Roman labels represent the weight of each hyperedge. The orange circle represents the initial hyperedge, and the green circle represents the newly created hyperedge. The purple solid lines and circles represent the 2nd-order and 1st-order hyperedges respectively. The dotted circles represent the higher-order moments. 
(a) A hypergraph $H$, corresponding to four modes and two hyperedges: one 2nd-order hyperedge $\lbrace A,D \rbrace$ with weight 2 and the other 3rd-order hyperedge $\lbrace A,B,C \rbrace$ with weight 1. On the corresponding quantum state, one can apply different Gaussian operations:
(b) Momentum displacement $Z^B(-2)$ on mode $B$ adds a hyperedge representing $\e^{-2 i\qhat_B}$ on it.
(c) Position displacement $X^A(1)$ on mode $A$ generates new hyperedges: a 1st-order hyperedge representing $\e^{-2 i \qhat_D}$ on $D$ and a 2nd-order hyperedge representing $\e^{-i \qhat_B\qhat_C}$ for $\{B,C\}$. (d) Position shearing $D_q^B(2)$ on mode $B$ adds a decoration representing $\e^{i \qhat_B^2}$, which is not part of the formalism. (e) Momentum shearing $D_p^A(1)$ on mode $A$ adds decorations, representing $\e^{i2 \qhat_D^2}$ and $\e^{i \qhat_B^2 \qhat_C^2/2}$, which fall outside the formalism, but also induces a valid new hyperedge representing $\e^{i 2\qhat_B\qhat_C \qhat_D}$ for $\{B,C,D\}$. (f) Squeezing $S^A(1)$ on mode $A$ changes the weights on all hyperedges adjacent to mode $A$. (g) Rotation $R^D(\pi)$ on mode $D$ changes the weights by inverting all signs on hyperedges adjacent to $D$. 
(h) Controlled-$Z$ gate $C_{A,D}(0.5)$ on modes $A$ and $D$ adjusts their weight by 0.5.
}
\label{fig:hgraph_ex2}
\end{figure}

For clarity, we give an example of a four-mode hypergraph state containing two hyperedges, $\ket{H} =\e^{i \qhat_{A} \qhat_{B} \qhat_C} \e^{i 2 \qhat_{A} \qhat_{D}} \ket{0}_p^{\otimes 4}$,
as shown in \cref{fig:hgraph_ex2}(a). One hyperedge $e_1=\lbrace A,B,C \rbrace$ has weight 
$t_{e_1}=1$ and the other hyperedge $e_2=\lbrace A,D \rbrace$ has weight $t_{e_2}=2$. 

In order to present the graphical framework more conveniently, we first introduce 
some basic rules of set operations. The hyperedge set $E_{a}$  is defined as the set of all hyperedges containing vertex $a$. 
The adjacency set $\AAA(a)$ of vertex $a \in V$ is defined as $\AAA(a)= \lbrace e \setminus \lbrace a \rbrace \mid e \in E_a \rbrace$. 
Considering the bijective mapping between hyperedge and weight sets, we define the union $(E_U; T_U) = (E_1; T_1) \cup (E_2; T_2)$ as a union of all tuples. For hyperedges appearing in both sets, we add up their weights. 
For the hypergraph in \cref{fig:hgraph_ex2}(a), the set of hyperedges containing mode $A$ is given by $E_A = \lbrace \lbrace A,B,C \rbrace, \lbrace A,D \rbrace \rbrace$ with associated weights $T_A = \{ 1, 2 \}$ and  $\mathcal{A}(A)=\lbrace \lbrace B,C \rbrace, \lbrace D \rbrace \rbrace$. If we unite $E_A$ and $E_B=\{\{A,B,C\}\}$ with $T_B = \{1 \}$, $(E_U; T_U)=( \lbrace \lbrace A,B,C \rbrace, \lbrace A,D \rbrace \rbrace; \{2,2\})$ is obtained.

Although not all quantum states can be transformed under the graphical rules, some states are able to be represented indirectly in the form of hypergraphs. For example, the state after applying the cubic phase gate $\e^{i \qhat^3}$ which contains a third-order moment term of one single-mode operator and which plays important roles in universal quantum computing \cite{PhysRevA.64.012310,Lloyd99quantumcomp,Kalajdzievski2019gatedecomposition}. We will demonstrate later that it can be represented as a sequence of operations similar to those of our formalism.

\section{Results} \label{sec:gaussopera}

In this section, we provide unified graphical rules of Gaussian unitary operations and Gaussian measurements for CV hypergraph states. We show the rules of single-mode Gaussian operations including displacement, shearing, squeezing, and rotation \cite{Gu_2009,Weedbrook_2012} and further discuss the CV controlled-$Z$ gate. Derivations are given in \cref{app:proofs}.
In principle, any multimode Gaussian operation can be represented as a concatenation of a beam splitter, single-mode squeezing, and phase shifts through the Bloch-Messiah decomposition \cite{PhysRevA.71.055801}, where the beam-splitter operation is equivalent to a controlled-$Z$ gate \cite{PhysRevResearch.6.023332}.
Note that some of the fundamental operations generate additional terms, which need to be taken into account when applying multiple operators consecutively.

\subsection{Displacement} \label{sec:displace}

The displacement operators include the momentum displacement $Z^a(s) = \e^{i s \qhat_a}$ and the position displacement $X^a(s) = \e^{-i s \phat_a}$, where $s$ represents the distance that the operator $\qhat_a$ or $\phat_a$ is moved in phase space \cite{PhysRevLett.88.097904}.
Thus, the momentum displacement operator $Z^a(s) = \e^{i s \qhat_a}$ on mode $a$ modifies the hyperedge set by $(E'; T') = (E;T) \cup (\lbrace a \rbrace; s )$. In other words, if mode $a$ has its own hyperedge initially, this operation will increase the weight by $s$. If not, a new hyperedge with the weight $s$ will be added to the system. \cref{fig:hgraph_ex2}(b) shows an example of performing operation $Z^B (-2)$ on the hypergraph state depicted in \cref{fig:hgraph_ex2}(a).

However, when performing the position displacement $X^a(s) = \e^{-i s \phat_a}$ to the hypergraph state, due to the incompatibility between two operators $\qhat_a$ and $\phat_a$, additional correlation terms will be generated  when they simultaneously act on the zero-momentum eigenstate $\ket{0}_p$. It influences hyperedges of the adjacency $\mathcal{A}(a)$ of mode $a$  such that $(E';T') = (E;T) \cup (\AAA(a); \{ \tilde{t} \})$, where $\tilde{t} \coloneqq - s \times t_{e \cup a}$. 
This means that the position displacement on mode $a$ will add or change hyperedges between its adjacent modes with the weight multiplied by $-s$. For example, when the position displacement  $X^A(1)$ is applied to mode $A$ of the hypergraph state corresponding to \cref{fig:hgraph_ex2}(a), two new hyperedges between vertices $\lbrace B,C \rbrace $ and $\lbrace D \rbrace$ of $\AAA(a)$ are created, as represented by the purple line and circle in \cref{fig:hgraph_ex2}(c).

Therefore, in a hypergraph state of $k$-th order, displacement operations can create new hyperedges of at most order $k-1$. 
Note that for CV graph states, the order of edges is $k=2$ and therefore displacement operations only cause local effects in a certain mode.

\subsection{Shearing} \label{sec:shear}

The shearing operator includes the momentum shearing $D_{p}^a(s) = \e^{i \frac{s}{2} \phat_a^2}$ and the position shearing $D_{q}^a(s) = \e^{i \frac{s}{2} \qhat_a^2}$, which shears the state with respect to the $\phat$ or $\qhat$-axis by a gradient of $s$. 
The shearing operator introduces an additional second-order moment term in the exponent, which does not occur in a standard hypergraph state, so this operation leaves the set of hypergraph states. This is in line with observations made on hypergraph states with qubits \cite{Gachechiladze17localcomplement}. 
For the position shearing operation, a term without effecting the hyperedge set of the state,
$D_{q}^a(s) \ket{H} = \e^{i \frac{s}{2} \qhat_a^2} \ket{H}$, will be created. For example, performing the position shearing $D_{q}^B(2)$ on mode $B$ will change the hypergraph into the one shown in \cref{fig:hgraph_ex2}(d). The dotted line indicates that there is a second-order moment on its own. We discuss how to deal with these extra decorations in \cref{app:proofs}.

Similarly, due to the incompatibility between two operators $\qhat_a$ and $\phat_a$, the momentum shearing operation $D_{p}^a(s)$ on mode $a$ will introduce new terms acting on the adjacency $\AAA(a)$. 
More precisely, we have $D_{p}^a(s) \ket{H}=e^{i \frac{s}{2}(\sum_{e \in \AAA(a)} t_{e \cup a}\qhat_{e})^2}\ket{H}$.  
In most cases, momentum shearing operation leads to a state which is not a hypergraph state. In the example of \cref{fig:hgraph_ex2}(a), the operation $D_p^A(1)$ on mode $A$ will introduce the remaining unitary $\e^{i (\qhat_B \qhat_C+2\qhat_D) ^2/2}$ to the original hypergraph state. Expanding the square term we get a forth-order moment term $\e^{i \qhat_B^2 \qhat_C^2/2}$, a second-order moment term $\e^{i 2 \qhat_D^2}$, and a hyperedge $\e^{i 2\qhat_B \qhat_C \qhat_D}$ among modes $\{B, C, D\}$, as shown in \cref{fig:hgraph_ex2}(e).

Therefore, in a hypergraph state of $k$-th order, a new hyperedge of $k$-th order can be generated to connect different modes by shearing operations. This effect can also link more pairs in the graph state. In principle, if a hypergraph state contains one $k$-th order hyperedge and another $l$-th order hyperedge which intersect in exactly one mode, shearing operations acting on this shared mode can expand the maximum order to $(k+l-2)$, offering a practical approach to generate more or higher-order hypergraph states with multimode nonlinearity using only local Gaussian operations.

\subsection{Squeezing} \label{sec:squeez}

Applying a single-mode squeezing operator $S^a(s) = \e^{-i \frac{s}{2} (\phat_a \qhat_a + \qhat_a \phat_a)}$ on mode $a$  multiplies the weight of all hyperedges linked to mode $a$ by an additional factor of $\e^{-s}$, while other hyperedges remain unchanged. 
That is, it updates the edge and weight sets by $(E';T') = (E;T) \cup (E_a; \{ \tilde{t} \})$, where $\tilde{t} \coloneqq (\e^{-s}-1) \times t_{e}$.

In \cref{fig:hgraph_ex2}(f), we show an example by squeezing mode $A$ with $S^A (1)$, hence, $t_{AD}=2\e^{-1}$ and $ t_{ABC}=\e^{-1}$.
Therefore, by using a squeezing operation, the value of weights in the hyperedges can be effectively proportionally adjusted, that is, the strength of the correlation between different modes can be changed. 

\subsection{Rotation}   \label{sec:rot}

Applying a local rotation $R_a(s) = \e^{i \frac{s}{2} (\qhat_a^2 + \phat_a^2)}$ in phase space usually results in the final state no longer being a standard hypergraph state. But in some special cases, $s = n \pi$, we still get a hypergraph state. 
The hypergraph keeps unchanged for $n$ is even, while the weights of hyperedges in $E_a$ will be negated for $n$ is odd. 
Particularly, the edge and weight sets are updated by $(E';T') = (E;T) \cup (E_a; \{ \tilde{t} \})$, where $\tilde{t} \coloneqq ((-1)^n-1) \times t_{e}$.

In \cref{fig:hgraph_ex2}(g), we show the effect of rotation $R^D (\pi)$ on mode $D$. This is consistent with the observations in the case of hypergraph states with qubits~\cite{comment}. 

For the general case $s \neq n \pi$, the specific form of the hypergraph state after rotation operation is given in \cref{app:proofs}. Since graph and hypergraph states originate from a series of logic gates acting on the zero-momentum eigenstates, but rotation operations may cause a shift to the zero-position eigenstate, some complicated correlation terms will be introduced by such a shift. Therefore, the use of rotation operations in the context of manipulating hypergraph states is not straightforward.
To avoid using rotations, we discuss how to replace the rotation operator by a sequence of displacement and shearing operators later in this section, as well as in \cref{app:proofs}.

\subsection{Cotrolled-Z gate} \label{sec:CZ}

It has been proven that any two-mode Gaussian operation can be decomposed into a CV controlled-$Z$ gate and a series of single-mode operations \cite{PhysRevA.71.055801,PhysRevResearch.6.023332}. 
So it suffices to focus on this two-mode Gaussian operation. A controlled-$Z$ gate $C_{i,j} (s)= \e^ {i s \qhat_i \qhat_j}$ can connect two modes $i$ and $j$ by an edge of strength $s$ or change its weight if edge $\{ i,j \}$ was there before. 
That is,  $(E';T') = (E;T) \cup (\{i,j\}; s)$.
In \cref{fig:hgraph_ex2}(h), we show the effect of $C_{A,D} (0.5)$ on modes $A$ and $D$. 

Therefore, the controlled-$Z$ gate provides a way to adjust the adjacency set of any vertex. By combining it with single-mode operations, we can create or delete hyperedges by Gaussian operations. 

\subsection{General Gaussian Operations}

In the previous subsections, we have introduced graphical rules for commonly used single-mode and two-mode Gaussian operations. Any other Gaussian operation can be decomposed into those operations. This can be seen by composing the following results: First of all, any Gaussian operation can be decomposed into first- and second-order operations \cite{arvind1995real}. It was shown in Ref.~\cite{PhysRevA.71.055801}, that second-order operations can be decomposed into a series of squeezing operations, beam splitter operations, and rotation operations. Furthermore, the effect of beam splitter operation can be decomposed into rotation, squeezing, and controlled-Z gate \cite{PhysRevResearch.6.023332,PhysRevA.99.022341}. 
In case that a rotation $R(s)$ with an angle $s$ different from $n \pi$, we can further decompose it into sheering and squeezing by $R(s)=D_p(t)S(r)D_q(t)$, where $t=\tan (s)$ and  $e^r=1/\cos (s)$. This is proven in \cref{app:proofs}. 
Therefore, any Gaussian operation can be written as a decomposition of operators discussed in \cref{sec:displace,sec:shear,sec:squeez,sec:rot,sec:CZ}. 

We have graphical rules for the building blocks of general Gaussian operations. However, if we apply multiple operators consecutively, we might have to deal with remaining unitaries. In \cref{app:proofs}, we show how those remaining unitaries commute with the set of discussed operators. 
By using the identities from \cref{app:proofs}, we can track the propagated remaining unitaries and defer them to the end.

\subsection{Gaussian Measurements}
 
Since a mode collapses after measurement, measuring $\dyad{m}_{q,a}$ will disconnect mode $a$ from the hypergraph and delete mode $a$ from all hyperedges which were adjacent to it before.
The hyperedge set and weight set change to
$(E';T') = ( E \setminus E_a; T ) \cup (\AAA(a); \{\bar{t}\})$, where the weights of all hyperedges in $\AAA(a)$ are $ \bar{t}\coloneqq m \times t_{e \cup a}$.

Measuring $\dyad{m}_{p,a}$ will also disconnect mode $a$ and modify hyperedges which were previously adjacent to mode $a$ by averaging over all possible weights. The generating state is given by
\begin{align}
\dyad{m}_{p,a} \ket{H} 
    =&\ket{m}_{p,a} \int_{-\infty}^{\infty} dx \e^{-i m x}  \prod_{e \in \AAA(a)} C_e(x t_{e \cup a})  \notag \\
    &\prod_{e' \in E_{\neg a}} C_{e'} (t_{e'}) \ket{0}_p^{\otimes \lvert V \rvert -1},
\end{align}
where $E_{\neg a} = \lbrace e   \mid e \in E, a \notin e \rbrace$ is the set of hyperedges without mode $a$.

\section{Discussion}

In this section, we discuss the experimental feasibility and possible applications of these graphical rules, and furthermore, we summarize our results.

\subsection{Experimental Feasibility} \label{sec:experimental}

Based on the graphical framework, it is possible to generate hypergraph states with richer structures from an initial 3rd-order hypergraph state using Gaussian operations. 
To realize the 3rd-order hypergraph state
$e^{i g \tau \qhat_A \qhat_B \qhat_C} \ket{0}_p^{\otimes 3}$, a Hamiltonian $H=g \qhat_A \qhat_B \qhat_C$ needs to be constructed, which acts on three zero-momentum eigenstates for the evolution time $\tau$.
Recently, a three-mode non-Gaussian state has been experimentally achieved via three-photon SPDC starting from a vacuum state in superconducting systems \cite{Chang20spntaneousdownconv}. 
When the state evolves from a squeezed state, the obtained state will approximate a hypergraph state as the squeezing level increases. 
Another interesting thing is finding a classification of hypergraph states which are equivalent up to Gaussian operations. Such a classification might further facilitate the preparation of hypergraph states in experimental contexts, commencing from a simply prepared state but with the same structure. It could be done similarly to the classification of local unitary or local Clifford equivalent qubit graph states \cite{vandenNest04graphicalClifforts,danielsen2005database12qubits,cabelloEntanglementEightqubitGraph2009,ji2010lulcconj,vandre2024distinguishinggraphstatesproperties, burchardt2025algorithmverifylocalequivalence, claudet2025decidinglocalunitaryequivalence}.
Related questions for CV states were studied in Refs.~\cite{Giedke_2014_LU_gaussian,Zhuang18resource,PhysRevX.8.041038}.

Moreover, it should be noted that the above derivations start from the hypergraph state with the limit of infinite squeezing. Considering the actual experimental realisations, it is anticipated that this graphical approach can be extended to encompass a broader class of approximate hypergraph states with finite squeezing, which fosters a deeper understanding of the underlying structure and dynamics of non-Gaussian states in CV systems. As was done for graph states \cite{Menicucci11graphrules,Kala:22}, we provide an example to address the effects of finite squeezing in \cref{sec:finite_squeez}, where the position displacement operation is applied.

\begin{figure}
    \centering
\includegraphics[width=0.45\textwidth]{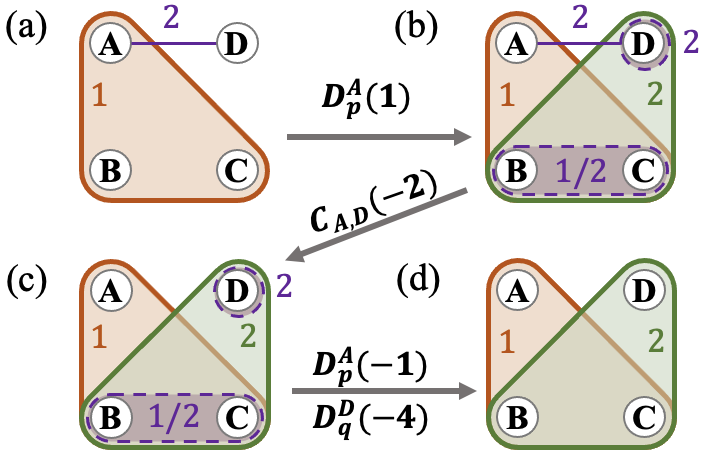}
    \caption{Generation of a hypergraph state with two 3rd-order hyperedges starting from a hypergraph state with only one 3rd-order hyperedge by Gaussian operations.  (a) The initial hypergraph state $\ket{H}$. (b) After performing momentum shearing on mode $A$, the desired green hyperedge of $\lbrace B,C,D \rbrace$ is generated. 
    (c) Removing the $\lbrace A,D \rbrace$ edge by performing a controlled-$Z$ gate $C_{A,D} (-2)$. (d) After performing the shearing operations on modes $A$ and $D$ respectively, the hypergraph state $\e^{i \qhat_A \qhat_B \qhat_C}\e^{i 2 \qhat_B \qhat_C \qhat_D}\ket{0}_p^{\otimes 4}$ is obtained.}
    \label{fig:example3edges}
\end{figure}

\subsection{Applications} 

The ability to generate non-Gaussian states with higher multimode nonlinearity is seen as a key resource in many applications. Directly executing non-Gaussian operations across multiple modes proves to be highly challenging in practice.
In \cref{fig:example3edges}, we show how to generate an additional 3rd-order hyperedge among modes $\{B, C, D\}$ from the initial hypergraph state as shown in \cref{fig:example3edges}(a). That is, performing a series of Gaussian operations to implement an equivalent result of a typical non-Gaussian operation, CV Toffoli gate, $C_{B,C,D}(2)=\e^{i 2 \qhat_{B} \qhat_{C}  \qhat_{D}}$, among them \cite{PhysRevResearch.6.023332}. 
More explicitly, we first use a local position shearing operation on mode $A$ to create a new 3rd-order hyperedge among its adjacency $\{B, C, D\}$, as shown in \cref{fig:example3edges}(b). This operation brings remaining unitaries (dotted circles) which can be further purified by Gaussian operations. We then perform a CV controlled-$Z$ gate between modes $A$ and $D$ to make them disconnected. Therefore, we apply two local shearing operations towards the opposite direction on modes $A$ and $D$ respectively, and then obtain the final four-mode hypergraph state with two 3rd-order hyperedges. 
In this way, we can directly determine the graphical evolution process of the state for on-demand generation of non-Gaussian quantum states, without complicated calculation. 
A further extension of this example is given in \cref{app:examples}.

\begin{figure}
	\centering
	\includegraphics[width=0.47\textwidth]{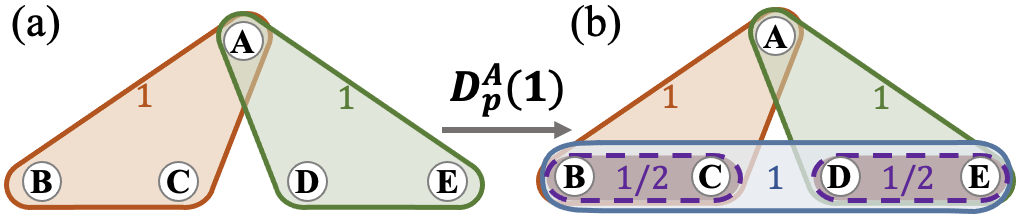}
	\caption{Generation of a hypergraph state with a 4th-order hyperedge out of a state with two 3rd-order hyperedges by  local shearing operation.
		(a) The initial hypergraph state $\e^{i \qhat_A \qhat_B \qhat_C}\e^{i \qhat_A \qhat_D \qhat_E}\ket{0}_p^{\otimes 5}$.
		(b) After performing momentum shearing on mode $A$, the desired blue hyperedge representing  $e^{i\qhat_B\qhat_C\qhat_D \qhat_E}$ is generated. Additionally, two purple dashed lines representing high-order moments, $e^{i\qhat_B^2 \qhat_C^2/2}$ and $e^{i\qhat_D^2 \qhat_E^2/2}$, are produced as remaining unitaries.
	}
	\label{fig:example4edges}
\end{figure}

Furthermore, it is possible to increase the cardinality of the graph by using the presented graphical framework. 
In \cref{fig:example4edges}, we show an example
where we turn a 3rd-order hypergraph state into a 4th-order hypergraph state under Gaussian operations.
The initial state has two 3rd-order hyperedges $\{A, B, C\}$ and $\{A, D, E\}$ which intersect in mode $A$. By applying a local position shearing operation on mode $A$, we obtain a four-mode nonlinearity process represented by a 4th-order hyperedge $\{B, C, D, E\}$, which is extremely difficult to achieve in practice.
Note that the resulting state as shown in \cref{fig:example4edges}(b) is not a standard hypergraph state, but we can remove these remaining unitaries in a similar way via auxiliary modes.

In addition to being able to prepare different types of non-Gaussian states, we demonstrate the possibility of generating the action of the cubic phase gate $\e^{i \qhat^3}$ \cite{PhysRevA.64.012310,Lloyd99quantumcomp} by using similar methods. 
Due to the commutative relation between operators $\qhat_j$ and $\phat_j$, we can obtain the equation $\e^{i \qhat_A \phat_C} \e^{i\qhat_A^2 \qhat_C}= \e^{i \qhat_A^3}\e^{i \qhat_A^2 \qhat_C} \e^{i \qhat_A \phat_C}$. Thus, the cubic phase gate can be decomposed by $\e^{i \qhat_A^3}=\e^{i \qhat_A \phat_C} \e^{i\qhat_A^2 \qhat_C} \e^{-i \qhat_A \phat_C} \e^{-i \qhat_A^2 \qhat_C}$. The Gaussian term $\e^{i \qhat_A \phat_C}$ can be  implemented through a phase-modulation controlled-$Z$ gate executed on modes $A$ and $C$. While the other term $\e^{i\qhat_A^2 \qhat_C}$ can be further decomposed in the same way $\e^{i \qhat_A^2 \qhat_C}=\e^{i \qhat_A \phat_B} \e^{i \qhat_A \qhat_B \qhat_C} \e^{-i \qhat_A \phat_B} \e^{-i \qhat_A \qhat_B \qhat_C}$, in which the term $\e^{i \qhat_A \qhat_B \qhat_C}$ is a CV Toffoli gate and associated to a 3rd-order hyperedge. When we apply the cubic phase gate on a proper hypergraph state, this Toffoli gate can be implemented like our first example. 
So, using the graphical framework, hypergraph states together with Gaussian operations results in a feasible protocol for the implementation of the cubic phase gate, thereby expanding the toolbox for its experimental realization.

\subsection{Summary}

In summary, we have extended the concept of weighted hypergraph states and developed a graphical framework tailored for CV non-Gaussian states. 
This framework starts from an initial hypergraph state, upon which the action of typical Gaussian operations and Gaussian measurements can be directly described.
Our graphical framework offers a visually intuitive way of manipulating CV non-Gaussian states, thereby presenting a valuable tool set that could significantly facilitate the design of quantum circuits in the realm of quantum computation. 

\section*{Acknowledgments}  
We thank Darren W.~Moore for discussions. This work was supported by the National Natural Science Foundation of China (No.~12125402, No.~12350006), the Innovation Program for Quantum Science and Technology (No.~2021ZD0301500), the National Cryptologic Science Fund of China (Grant No.~2025NCSF02048), Beijing Natural Science Foundation (Grant No.~Z240007), the Deutsche Forschungsgemeinschaft  (DFG, German Research Foundation, project numbers 447948357 and 440958198), the Sino-German Center for Research Promotion (Project M-0294), the German Ministry of Education and Research (Project QuKuK, BMBF Grant No.\ 16KIS1618K), the Stiftung der Deutschen Wirtschaft, 
and the Austrian Science Fund (FWF) project quantA [10.55776/COE1]. For Open Access purposes, the author has applied a CC BY public copyright license to any author accepted manuscript version arising from this submission.

\onecolumn
\appendix

\section{Derivations of the graphical framework}\label{app:proofs}

In the following, we provide the derivation details of the results presented above. 
Since the operations $C_e$ commute for all hyperedges $e \in E$, the identity 
\begin{align}
\prod_{e \in E} C_e(t_e) = \prod_{e \in E} \e^{i t_e \qhat_e} = \e^{i \sum_{e \in E}  t_e \qhat_e}
\end{align}
holds.
We can divide the set by a vertex $a \in V$ into disjoint subsets $E = E_a \dot{\bigcup} E_{\neg a}$, where $E_a = \lbrace e   \mid e \in E, a \in e \rbrace$ is the set of hyperedges containing $a$, and $E_{\neg a} = \lbrace e   \mid e \in E, a \notin e \rbrace$ is the set of hyperedges not containing $a$.
We therefore have 
\begin{align}
\nonumber
\prod_{e \in E} C_e(t_e) 
&= \e^{i \sum_{e \in E_a}  t_e \qhat_e} \prod_{e' \in E_{\neg a}} C_{e'} (t_{e'}) \\ 
&= \e^{i \qhat_a h(a)} \prod_{e' \in E_{\neg a}} C_{e'} (t_{e'}),
\end{align}
where we define
$h(a) \coloneqq \sum_{e \in \AAA(a)}  t_{e \cup a} \qhat_{e}$. Note that by definition $E_a = \lbrace e \cup a \vert e \in \AAA(a) \rbrace$.

\subsection{Displacement in momentum}
The momentum displacement operator $Z^a(s) = \e^{i s \qhat_a}$ modifies the hyperedge set by $(E;t) \rightarrow (E;t) \cup (\lbrace a \rbrace; s )$. That is, it adds the hyperedge $\lbrace a \rbrace$ with the weight $s$ if there was no such hyperedge before or changes the weight by $s$.

For $\lbrace a \rbrace \in E$:
\begin{align}
\nonumber
    Z^a(s) \ket{H} &=  \e^{i s \qhat_a} \e^{i t_a \qhat_a}  \prod_{e \in E \setminus \lbrace a \rbrace} C_e(t_e) \ket{0}^{\otimes \vert V \vert} \\
    &=  C_a(s+t_a)  \prod_{e \in E \setminus\lbrace a \rbrace} C_e(t_e) \ket{0}^{\otimes \vert V \vert}. 
\end{align}

The case $\lbrace a \rbrace \notin E$ directly follows by setting $t_a=0$ in the above calculation.

\subsection{Displacement in position}
Applying the position displacement operator $X(s)$ on the mode $a$ modifies the hyperedges in the adjacency $\AAA(a)$ of $a$ such that $(E';t') = (E;t) \cup (\AAA(a); \tilde{t})$, where $\tilde{t}_e \coloneqq -s \times t_{e \cup a}$. In other words, it adds or updates the hyperedges in the adjacency $\AAA(a)$ of $a$ and multiplies the weight of the hyperedge containing $a$ by $s$:
\begin{align}
\nonumber
    X^a(s) \ket{H} &= \e^{-i s \phat_a} \prod_{e \in E} C_e(t_e) \ket{0}_p^{\otimes \lvert V \rvert } \\
    \nonumber
    &= \e^{i \qhat_a h(a)} \e^{-i s \phat_a} \e^{-i s h(a)} \prod_{e' \in E_{\neg a}} C_{e'} (t_{e'}) \ket{0}_p^{\otimes \lvert V \rvert } \\
    &= \prod_{e \in E} C_e(t_e)  \prod_{e'' \in \AAA(a)} C_{e''}(-s \dot t_{e'' \cup a}) \ket{0}_p^{\otimes \lvert V \rvert }.
\end{align}

\subsection{Shearing}
Applying shearing in position is trivial. This expression 
\begin{align}
    D_{q}^a(s) \ket{H} 
    &= \e^{i \frac{s}{2} \qhat_a^2}  \prod_{e \in E} C_{e} (t_{e}) \ket{0}_p^{\otimes \lvert V \rvert } 
\end{align}
cannot be simplified.

Applying shearing in  momentum gives us
\begin{align}
\nonumber
    D_{p}^a&(s) \ket{H} 
    = \e^{i \frac{s}{2} \phat_a^2} \e^{i \qhat_a h(a)} \prod_{e' \in E_{\neg a}} C_{e'} (t_{e'}) \ket{0}_p^{\otimes \lvert V \rvert } \\
    \nonumber
    &= \e^{i \qhat_a h(a)} \e^{is h(a) \phat_a} \e^{i \frac{s}{2} h(a)^2 }  \prod_{e' \in E_{\neg a}} C_{e'} (t_{e'})  \ket{0}_p^{\otimes \lvert V \rvert } \\
    \nonumber
    &=  \e^{i \frac{s}{2} h(a)^2}  \prod_{e \in E} C_{e} (t_{e})  \ket{0}_p^{\otimes \lvert V \rvert } \\
    \nonumber
    &=  \e^{i \frac{s}{2} (\sum_{e \in \AAA(a)}  t_{e \cup a} \qhat_{e})^2}  \ket{H} \\
    &= \prod_{e \in \AAA(a)}  \e^{i \frac{s}{2}   t_{e \cup a}^2 \qhat_{e}^2}
    \prod_{e, e' \in \AAA(a), e \neq e'} \e^{i s \cdot  t_{e \cup a} t_{e' \cup a} \qhat_{e} \qhat_{e'}}
    \ket{H},
\end{align}
where we have used used the identity 
$\left[ \phat_a^2, \qhat_a  \right] = - 2 i \phat_a$.

\subsection{One-mode squeezing}
Applying a one-mode squeezing operation $S^a(s) = \e^{-i \frac{s}{2} (\phat_a \qhat_a + \qhat_a \phat_a)} = \e^{-\frac{s}{2}} \e^{-i s \qhat_a \phat_a}$ on the mode $a$ changes the weight of all hyperedges containing $a$ by a factor of $\e^{-s}$:
\begin{align}
\nonumber
    S^a(s) \ket{H} 
   & = e^{-\frac{s}{2}} \e^{-i s \qhat_a \phat_a } \e^{i \qhat_a h(a)} \prod_{e' \in E_{\neg a}} C_{e'} (t_{e'}) \ket{0}_p^{\otimes \lvert V \rvert } \\
    \nonumber
    &= e^{-\frac{s}{2}} \e^{i \e^{-s} \qhat_a h(a)}  \prod_{e' \in E_{\neg a}} C_{e'} (t_{e'}) \ket{0}_p^{\otimes \lvert V \rvert } \\
    &= e^{-\frac{s}{2}} \prod_{e \in E_a} C_e(\e^{-s} t_e)   \prod_{e' \in E_{\neg a}} C_{e'} (t_{e'}) \ket{0}_p^{\otimes \lvert V \rvert } .
\end{align}
We have used the Baker-Campbell-Hausdorff identity
\begin{align}
    \e^X e^Y = e^{(Y + [X,Y] + \frac{1}{2!}[X,[X,Y]] + \frac{1}{3!}[X,[X,[X,Y]]] + ...)} e^X, \label{eq:BCH}
\end{align}
with $X = -i s \qhat_a \phat_a$ and $Y = i \qhat_a h(a)$.

\subsection{Rotation}  \label{sec:proof_rot}
Applying a rotation operator in general gives us a state which is not in the form of a hypergraph state. The general form is given by:
\begin{align}
\nonumber
    R^a(s) &\ket{H} 
    = \e^{i \frac{s}{2} (\qhat_a^2 + \phat_a^2)} \e^{i \qhat_a h(a)} \prod_{e' \in E_{\neg a}} C_{e'} (t_{e'}) \ket{0}_p^{\otimes \lvert V \rvert } \\
    \nonumber
    &= \e^{i   h(a) ( \cos(s) \qhat_a + \sin(s) \phat_a)} \e^{i \frac{s}{2} (\qhat_a^2 + \phat_a^2)}
    \nonumber
    \prod_{e' \in E_{\neg a}} C_{e'} (t_{e'}) \ket{0}_p^{\otimes \lvert V \rvert } \\
    &= \e^{i \qhat_a h(a) \cos(s)} \e^{i \phat_a h(a) \sin(s)} \e^{i \frac{1}{4} h(a)^2 \sin(2s)}
    \prod_{e' \in E_{\neg a}} C_{e'} (t_{e'}) \e^{i \frac{s}{2} (\qhat_a^2 + \phat_a^2)} \ket{0}_p^{\otimes \lvert V \rvert } \\
    &= \e^{i \qhat_a h(a) \cos(s)} \e^{i \phat_a h(a) \sin(s)} \e^{i \frac{1}{4} h(a)^2 \sin(2s)}
    \nonumber
    \prod_{e' \in E_{\neg a}} C_{e'} (t_{e'}) \ket{0}_p^{\otimes \lvert V \rvert -1} (R^a(s)\ket{0}_{p,a}
    ),
\end{align}
where $R^a(s)\ket{0}_{p} = -\sin(s)\ket{0}_q + \cos(s) \ket{0}_p$.
We used the Baker-Campbell-Hausdorff identity shown in \cref{eq:BCH}
with $X = i \frac{s}{2} (\qhat^2 + \phat^2)$ and $Y = i \qhat h(a)$.

Another special case is the Fourier transform $F = R(\frac{\pi}{2}) $, which swaps $\qhat$ with $\phat$, and $\phat$ with $-\qhat$.

The rotation operator can be decomposed into a sequence of squeezing and shearing operators. 
 In the following, we show that $R(s) = D_p(t)S(r)D_q(t)$ for $t=\tan (s)$ and  $e^r=1/\cos (s)$ for rotations $s \neq \frac{(2n+1)\pi}{2}$, where $n \in \Nd$.
The rotation, squeezing, and shearing operators in the phase space notation are given by 
\begin{align}
    D_q(t) &=
    \begin{pmatrix}
      1 & 0 \\
      t &  1 
    \end{pmatrix},
  \quad
 & S(r)  &=
    \begin{pmatrix}
      e^{r} & 0      \\
      0      & e^{-r}
    \end{pmatrix}, \notag \\
D_p(t)  &=
    \begin{pmatrix}
      1 & -t \\
      0 & 1
    \end{pmatrix},
  \quad
  &R(s) &=
    \begin{pmatrix}
      \cos (s) & -\sin (s) \\
      \sin (s) &  \cos (s)
    \end{pmatrix}.
\end{align}
It can be checked that $R(s) = D_p(t)S(r)D_q(t)$ holds for $t=\tan (s)$ and  $e^r=1/\cos (s)$.

\subsection{Controlled-$Z$ gate} 
The effect of application of the  controlled-$Z$ $C_{a,b} (s)= \e^ {i s \qhat_a \qhat_b}$ gate on two modes $a, b$ directly follows from the definition of hypergraph states. If the edge $\{a,b\} \in E$ was part of the edge set, its weight gets modified, since 
\begin{align}
    C_{\{a,b\}}(s) C_{\{a,b\}}(t_{\{a,b\}}) = C_{\{a,b\}}(s + t_{\{a,b\}})
\end{align}
holds. Otherwise, if $\{a,b\} \notin E$ it adds the edge $\{a,b\}$ with weight $s$ to the edge set. 

\subsection{General Gaussian Operations}

As discussed in the main text, general Gaussian operations can be decomposed into a sequence of the discussed operations \cite{arvind1995real, PhysRevA.71.055801, PhysRevResearch.6.023332,PhysRevA.99.022341}. In addition to the decompositions mentioned in the literature, we prove  the decomposition of the rotation operator in terms of squeezing and shearing operators in \cref{sec:proof_rot}.

Not all the discussed operators map hypergraph states to hypergraph states. The operations shearing and rotation map hypergaph states to hypergraph states up to remaining unitaries. The remaining unitary of a hypergraph state associated to the hypergraph $H = (V,E)$ are of the form $e^{if(\phat_1,\qhat_1, \dots, \phat_{\lvert V \rvert},\qhat_{\lvert V \rvert})}$, where $f(\phat_1,\qhat_1, \dots, \phat_{\lvert V \rvert},\qhat_{\lvert V \rvert})$
is a polynomial function of momentum and position operators of the modes corresponding to the vertex set $V$. In this section we show how the remaining unitaries in the most general form commute with the operators. We explicitly derive the relation for the shearing operator $D_q(s)$, the other relations are derived equivalently.

For computing the commutation relation of the shearing operator $D_q(s)$, we use the identities
\begin{align}
  D_q(s)\qhat D_q(-s)&=\qhat,  
  &D_q(s)\qhat^nD_q(-s)&=(\qhat )^n, \notag \\
  D_q(s)\phat D_q(-s)&=\phat -s \qhat, 
  &D_q(s)\phat^nD_q(-s)&=(\phat -s\qhat)^n.
\end{align}
Furthermore, we have
\begin{align}
    D_q(s)f(\phat, \qhat )D_q(-s) = f(D_q(s)\phat D_q(-s), D_q(s)\qhat D_q(-s)).
\end{align}
This can be seen by inserting identities $\1 = D_q(-s) D_q(s)$ between all products of operators in the function. Using the Taylor expansion, we get
\begin{align}
    D_q(s)e^{if(\phat, \qhat )}D_q(-s) = e^{if(D_q(s)\phat D_q(-s), D_q(s)\qhat D_q(-s))}.
\end{align}
For a multimode remaining unitary, we therefore have
\begin{align}
    D_q^a(s)e^{if(\phat_1,\qhat_1, \dots, \phat_{\lvert V \rvert},\qhat_{\lvert V \rvert})}=e^{if(p_1,q_1,...p_a-s q_a ,q_a ,...p_{\lvert V \rvert},q_{\lvert V \rvert})} D_q^a(s).
\end{align}

The commutation relations of other operators from \cref{sec:gaussopera} and the remaining unitary term are listed below.
\begin{align}
    X^a(s)e^{if(\phat_1,\qhat_1, \dots, \phat_{\lvert V \rvert},\qhat_{\lvert V \rvert})}
    &=e^{if(\phat_1,\qhat_1,\dots ,\phat_a,\qhat_a+s,\dots ,\phat_{\lvert V \rvert},\qhat_{\lvert V \rvert})} X^a(s), \nonumber\\
    Z^a(s)e^{if(\phat_1,\qhat_1, \dots, \phat_{\lvert V \rvert},\qhat_{\lvert V \rvert})}
    &=e^{if(\phat_1,\qhat_1,\dots ,\phat_a-s,\qhat_a,\dots ,\phat_{\lvert V \rvert},\qhat_{\lvert V \rvert})}Z^a(s), \nonumber \\
    D_p^a(s)e^{if(\phat_1,\qhat_1, \dots, \phat_{\lvert V \rvert},\qhat_{\lvert V \rvert})}
    &=e^{if(\phat_1,\qhat_1,\dots ,\phat_a,\qhat_a+s\phat_a,\dots ,\phat_{\lvert V \rvert},\qhat_{\lvert V \rvert})} D_p^a(s),  \nonumber\\
    D_q^a(s)e^{if(\phat_1,\qhat_1, \dots, \phat_{\lvert V \rvert},\qhat_{\lvert V \rvert})}
    &=e^{if(\phat_1,\qhat_1,\dots ,\phat_a-s\qhat_a,\qhat_a,\dots ,\phat_{\lvert V \rvert},\qhat_{\lvert V \rvert})} D_q^a(s),  \nonumber\\
    S^a(s)e^{if(\phat_1,\qhat_1, \dots, \phat_{\lvert V \rvert},\qhat_{\lvert V \rvert})}
    &=e^{if(\phat_1,\qhat_1,\dots ,e^{-s} \phat_a ,e^{s} \qhat_a ,\dots ,\phat_{\lvert V \rvert},\qhat_{\lvert V \rvert})} S^a(s),  \nonumber\\
    C_{a,b}(s) e^{if(\phat_1,\qhat_1, \dots, \phat_{\lvert V \rvert},\qhat_{\lvert V \rvert})}
    &=e^{if(\phat_1,\qhat_1, \dots,\phat_a-\qhat_b,\dots,\phat_b-\qhat_a,\dots, \phat_{\lvert V \rvert},\qhat_{\lvert V \rvert})}C_{a,b}(s).
\end{align}
By using these identities, we can track the propagated remaining unitaries and defer them to the end.

\subsection{Measurement in $\qhat$ basis} 
Measuring $\dyad{m}_{q,a}$  disconnects mode $a$ from the hypergraph by disconnecting hyperedges containing $a$. The hyperedge set and weights  get updated as
$E' = ( E \setminus E_a ) \cup \AAA(a) $ and for $e \in E'$
\begin{align}
    t'_e = 
    \begin{cases}
        m t_{e \cup a} & \text{if } e \in \AAA(a), \\
        t_e  & \text{else} .
    \end{cases}
\end{align}
In detail we have
\begin{align}
\nonumber
    \dyad{m}_{q,a} \ket{H} 
    &= \dyad{m}_{q,a}  \e^{i \qhat_a h(a)} \prod_{e' \in E_{\neg a}} C_{e'} (t_{e'}) \ket{0}_p^{\otimes \lvert V \rvert } \\
    \nonumber
    &= \dyad{m}_{q,a}  \e^{i m h(a)} \prod_{e' \in E_{\neg a}} C_{e'} (t_{e'}) \ket{0}_p^{\otimes \lvert V \rvert } \\
    &= \ket{m}_{q,a}  \prod_{e \in \AAA(a)} C_e(m t_{e \cup a})
    \prod_{e' \in E_{\neg a}} C_{e'} (t_{e'})  \ket{0}_p^{\otimes \lvert V \rvert -1} .
\end{align}

\subsection{Measurement in $\phat$ basis} 
Measuring $\dyad{m}_{p,a}$  disconnects mode $a$ from the hypergraph by disconnecting hyperedges containing $a$ and creates several hyperedges in a superposition of all possible weights, with a phase determined by the weight and the measurement result $m$.
The hyperedge set and weight set get updated as
$E' = ( E \setminus E_a ) \cup \AAA(a) $ and for $e \in E'$
\begin{align}
    t'(e) = 
    \begin{cases}
        x t(e \cup a) & \text{if } e \in \AAA(a), \\
        t(e)  & \text{else} .
    \end{cases}
\end{align}
In detail, we have
\begin{align}
\nonumber
    \dyad{m}_{p,a} \ket{H} 
    &= \dyad{m}_{p,a}  \e^{i \qhat_a h(a)} \prod_{e' \in E_{\neg a}} C_{e'} (t_{e'}) \ket{0}_p^{\otimes \lvert V \rvert } \\
    \nonumber
    &=\int_{-\infty}^{\infty} dx \dyad{m}_{p,a}  \e^{i x h(a)} \dyad{x}_{q,a} 
    \prod_{e' \in E_{\neg a}} C_{e'} (t_{e'}) \ket{0}_p^{\otimes \lvert V \rvert } \\
    &=\ket{m}_{p,a} \int_{-\infty}^{\infty} dx   \e^{i x h(a)} \e^{-i m x} 
    \prod_{e' \in E_{\neg a}} C_{e'} (t_{e'}) \ket{0}_p^{\otimes \lvert V \rvert -1}. 
\end{align}
The integral cannot be further simplified, so it is not a hypergraph state after measuring $\phat$.

\section{Effects of Finite Squeezing} \label{sec:finite_squeez}

According to the hypergraph state definition defined before, every vertex represents a infinitely squeezed zero-momentum eigenstate $\ket{0}_p = \lim_{r \rightarrow \infty} S(-r) \ket{0}$, where $\ket{0}$ is the vacuum state. In experimental realisations, these states have to be approximated by finitely squeezed states $\ket{0}_p \approx S(-r) \ket{0}$ for some large $r \in \mathbb{R}$.
The hypergraph state $\ket{H_{\text{ideal}}}=\e^{i \qhat_{A} \qhat_{B} \qhat_{C}}\ket{000}_p$ corresponding to the three vertex hypergraph $H$ with a single hyperedge $\{A,B,C \}$ is therefore approximated by $\ket{H_{\text{approx}}}=\e^{i \qhat_{A} \qhat_{B} \qhat_{C}} S_A(-r_A) S_B(-r_B) S_C(-r_C)\ket{000}$, where $r_A, r_B, r_C \in \mathbb{R}$.
In the derivation of the graphical rules, we make use of the identity $\e^{i f(\qhat,\phat) \phat} \ket{0}_p = \ket{0}_p$ which holds for all functions $f(\qhat,\phat)$ of $\qhat$ and $\phat$. However, in the case of finite squeezing, terms of the form $\e^{i f(\qhat,\phat) \phat}$ do not vanish. 
For example, applying the position displacement operator $X^A(s)$  on the two states $\ket{H_{\text{ideal}}}$ and $\ket{H_{\text{approx}}}$ leads to 
\begin{align}
     X^A(s) \ket{H_{\text{ideal}}} =& \e^{i \qhat_{A} \qhat_{B} \qhat_{C}} \e^{-i s \qhat_{B} \qhat_{C}} \ket{000}_p, \nonumber\\
     X^A(s) \ket{H_{\text{approx}}} = &\e^{i \qhat_{A} \qhat_{B} \qhat_{C}} \e^{-i s \qhat_{B} \qhat_{C}}\e^{-is \hat{p}_A}
    S_A(-r_A) S_B(-r_B) S_C(-r_C)\ket{000}. 
\end{align}
We can predict the state $\ket{H_{\text{ideal}}}$ by
\begin{align}
    \ket{H_{\text{predict}}}=&\e^{i \qhat_{A} \qhat_{B} \qhat_{C}}\e^{-i s \qhat_{B} \qhat_{C}} 
    S_A(-r_A) S_B(-r_B) S_C(-r_C)\ket{000}.
\end{align}
The fidelity of $ \ket{H_{\text{predict}}}$ and 
the actually displaced state $X^A(s) \ket{H_{\text{approx}}}$ is given by $\e^{-\frac{1}{2}s^2 \e^{-2r_A}}$, which tends to 1 for a large squeezing parameter $r_A$. 
Here, we assume that the states can be squeezed such that the fidelity between the actually prepared state and the state computed using the graphical rules from this paper is sufficiently high. 
The fidelity for other operations can be computed equivalently.

\section{Further Extension of the Example Given in \cref{fig:example3edges}} \label{app:examples}

\begin{figure}
    \centering
    \includegraphics[width=0.9\textwidth]{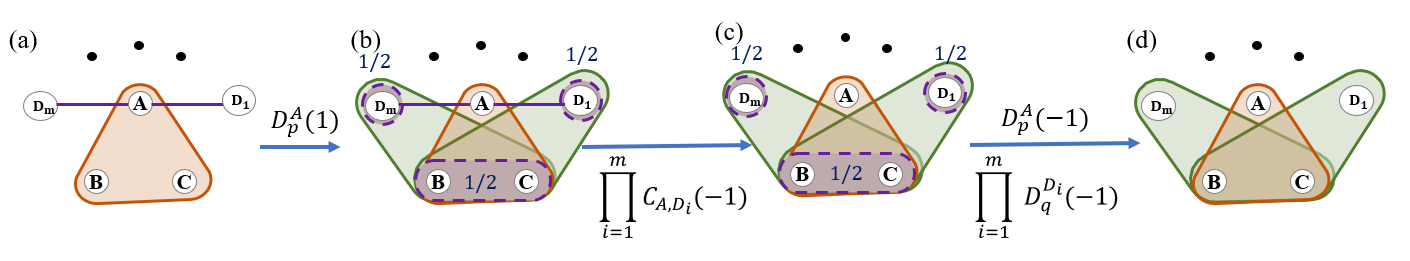}
    \caption{
    Generation of a state with $m$ 3rd-order hyperedges. (a) The initial hypergraph state with only one 3rd-order hyperedge. (b) After performing momentum shearing on mode $A$, the desired purple hyperedges of $\lbrace B,C,D_1 \rbrace$,\dots,$\lbrace B,C,D_m \rbrace$ are generated. Additionally, $\e^{i \frac{1}{2} \qhat_{D_1}^2}$,\dots, $\e^{i \frac{1}{2} \qhat_{D_m}^2}$, and $e^{i\frac{1}{2}\qhat_B^2 \qhat_C^2}$, are produced as remaining unitaries.
    (c) Removing the $\lbrace A,D_1 \rbrace$, \dots, $\lbrace A,D_m \rbrace$ edges by performing a series of two-mode controlled-$Z$ gates, $C_{A,D_1} (-1)$, \dots, $C_{A,D_m} (-1)$. 
    (d) After performing the shearing operations towards the opposite direction on modes $A$ and $D_1$, \dots, $D_m$ respectively, the new hypergraph state $\e^{i \qhat_A \qhat_B \qhat_C}\e^{i \qhat_B \qhat_C \qhat_{D_1}}$ \dots $\e^{i \qhat_B \qhat_C \qhat_{D_m}}\ket{0}_p^{\otimes {m+3}}$ is obtained. 
    }
    \label{fig:example_multiple3edges}
\end{figure}

In \cref{fig:example3edges}, we couple mode $D$ with mode $A$ and use the graphical rules to show that it generates a new hyperedge of order 3 between modes $B$, $C$, and $D$. An extended example is shown in \cref{fig:example_multiple3edges}. We couple multiple modes $D_1$, \dots, $D_m$ with mode $A$ and perform the same sequence of operations which has the same effect than applying $m$ Toffoli gates on the initial state.
We can see that, after performing the shearing operations on mode $A$ and a series of controlled-$Z$ gates as well as the shearing operations towards the opposite direction on modes $A$ and $D_1$, \dots, $D_m$ respectively, the new hypergraph state contains $m$ new hyperedges.
Instead of applying a non-Gaussian Toffoli gate on a certain hypergraph state, we can apply a sequence of Gaussian operations on the same state which has the same effect.

\bibliographystyle{quantum}
\bibliography{references}

\begin{thebibliography}{10}
\bibitem{comment}
A $\frac{\pi}{2}$ rotation is analogous to the application of the Hadamard gate
  in the qubit domain which in general leads out of the hypergraph formalism
  but specific cases are known where it results in a hypergraph state again.
\end{thebibliography}

\end{document}